\documentclass[cits]{PoS}

\title{The Brera Multi-scale Wavelet {\it Chandra} Survey. I. Serendipitous source catalogue}

\ShortTitle{The Brera Multi-scale Wavelet {\it Chandra} Survey}

\author{ \speaker{P.\ Romano},$^a$ S.\ Campana,$^b$ R.P.\ Mignani,$^c$ 
A.\ Moretti,$^b$ M.\ Mottini,$^{d}$ M.R.\ Panzera,$^b$  G.~Tagliaferri$^b$ \\
\llap{$^a$}INAF, Istituto di Astrofisica Spaziale e Fisica Cosmica, \\
         Via U.\ La Malfa 153, I-90146 Palermo, Italy\\
\llap{$^b$}INAF, Osservatorio Astronomico di Brera, \\ 
         Via E.\ Bianchi 46, I-23807 Merate, Italy\\
\llap{$^c$}Mullard Space Science Laboratory, University College London, \\
     Holmbury St.\ Mary, Dorking, Surrey, RH5 6NT, UK\\
\llap{$^d$}European Southern Observatory, \\
Schwarzschild Stra\ss e 2, 85740 Garching bei M\"unchen, Germany\\
E-mail: \email{romano@ifc.inaf.it}
}

\abstract{We present the Brera Multi-scale Wavelet {\it Chandra}  (BMW-{\it Chandra}) source catalogue 
drawn from essentially all {\it Chandra}   ACIS-I pointed observations with an
exposure time in excess of 10\,ks public as of March 2003 (136 observations).
Using the wavelet detection algorithm developed by Lazzati et al.\ (1999) and
Campana et al.\ (1999), which can characterise both point-like and extended
sources, we identified 21325 sources. 
Among them, 16758 are serendipitous, i.e.\ not associated with the targets 
of the pointings. 
This makes our catalogue the largest compilation of {\it Chandra} sources to date. 
The 0.5--10\,keV absorption corrected fluxes of these sources 
range from $\sim 3\times 10^{-16}$ to $9\times10^{-12}$ erg cm$^{-2}$ s$^{-1}$ with a median of 
$7\times 10^{-15}$ erg cm$^{-2}$ s$^{-1}$.
The catalogue consists of count rates and relative errors in three energy bands
(total, 0.5--7\,keV; soft, 0.5--2\,keV; and hard, 2--7\,keV), where
the detection was performed, and 
source positions relative to the highest signal-to-noise detection among the three bands. 
The wavelet algorithm also provides an estimate of the extension of 
the source.
We include information drawn from the headers of the original files, as well, and
extracted source counts in four additional energy bands, 
SB1 (0.5--1\,keV), SB2 (1--2\,keV), HB1 (2--4\,keV), and HB2
(4--7\,keV).
We computed the sky coverage for the full catalogue and for a subset
at high Galactic latitude ($\mid b \mid \, > 20^{\circ}$).
Our sky coverage in the soft band (0.5--2\,keV, S/N =3)  is 
$\sim 8$ deg$^2$ at a limiting flux of $\sim 10^{-13}$  erg cm$^{-2}$ s$^{-1}$,
and $\sim 2$ deg$^2$ at a limiting flux of $\sim 10^{-15}$ erg
cm$^{-2}$ s$^{-1}$.
}

\FullConference{7th INTEGRAL Workshop\\
		 September 8-11 2008\\
		 Copenhagen, Denmark}

\begin{document}

\section{Introduction}

The Brera Multi-scale Wavelet (BMW, \cite{Lazzatiea99,Campanaea99}) 
algorithm, which was developed to analyse {\it ROSAT} High Resolution Imager 
(HRI) images \cite{Panzeraea03}, was modified to support the 
analysis of {\it Chandra} Advanced CCD Imaging Spectrometer (ACIS) images 
\cite{Morettiea02}, and subsequently led to interesting results on the nature of 
the cosmic X-ray background \cite{Campanaea01}. 
Differently from other WT-based algorithms, the BMW automatically characterises 
each source through a multi-source $\chi^2$ fitting with respect to a Gaussian model 
in the wavelet space, and has therefore proven to perform well in 
crowded fields and in conditions of very low background \cite{Lazzatiea99}.
Given the reliability and versatility of the BMW,  we decided 
to apply it to a large sample of {\it Chandra} ACIS-I  images, 
to take full advantage of  the superb spatial resolution 
of {\it Chandra}  [$\sim 0.5$'' point-spread function (PSF) on-axis]. 
We thus produced the Brera Multi-scale Wavelet {\it Chandra} Survey
\cite{Romano2008}
and here we  present a pre-release of this catalogue,  
which is based on a subset of the whole  {\it Chandra} ACIS observations dataset,
roughly  corresponding to  the first  three years  of  operations.  
Our catalogue provides source positions, count rates,
extensions and relative errors.

\section{Method}

\subsection{Sample selection}

\begin{figure}
\vspace{-2truecm}
\includegraphics[width=1.0\textwidth,angle=0]{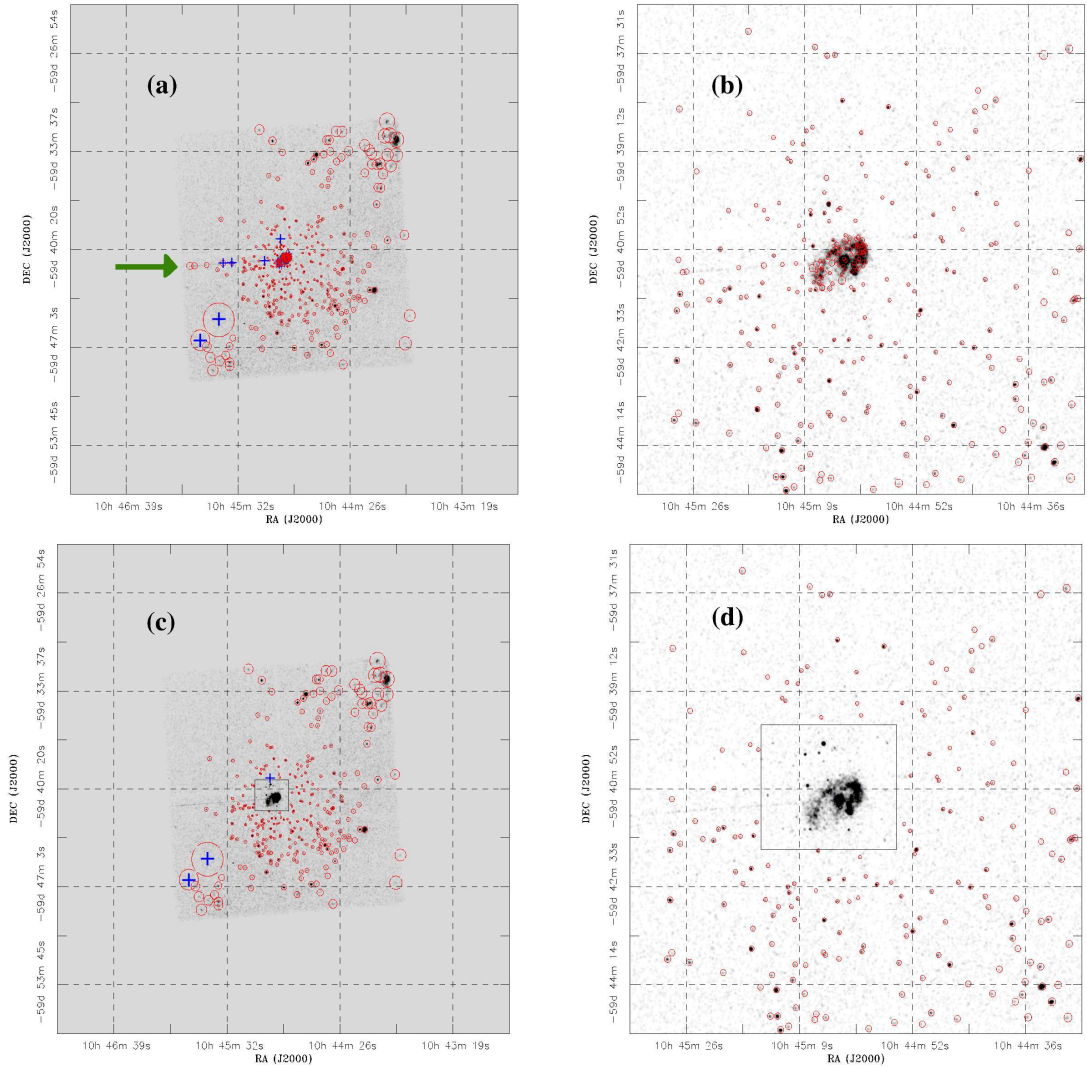}
\vspace{-3.5truecm}
\caption{Example of detection. {\bf (a)} The $\eta$ Carinae full field at half resolution. 
		Note the complicated extended structure at the centre and 
		the spurious detections along a readout streak (green arrow). 
		{\bf (b)} Central portion of the field at full resolution. 
		Crosses mark sources that the detection algorithm classifies as extended 
		(e.g.\ left-bottom corner and along readout streak).
		{\bf (c)} Example of manual cleaning. 
		The spurious sources along the readout streak were eliminated and {\bf (d)} 
		the sources in the central portion of the image 
		(contained within the box and not shown) were flagged for later inspection.}
\label{data}
\end{figure}

We chose the {\it Chandra} fields which 
maximised the sky area not occupied by the pointed targets, that is the
fields where the original PI was interested in a single, possibly 
point-like object centred in the field. Our criteria  were the following:
\begin{enumerate}
\item 	All ACIS-I [no grating, no High Resolution Camera (HRC) fields 
	and in Timed Exposure mode] 
	fields with exposure time in excess of 10\,ks available by 2003 March  
	were considered.
	Data from all four front-illuminated (FI) CCDs (I0, I1, I2, I3) were used. 
\item 	We excluded fields dominated by extended sources   
	[covering more than 1/9 of the field of view (FOV)]. 
\item 	We excluded planet observations  and supernova remnant observations.
\item 	We also excluded fields with bright point-like or high-surface 
        brightness extended sources. 
\item 	We put no limit on Galactic latitude, but  
	we selected sub-samples based on latitude at a later time. 
\end{enumerate}
The exclusion of bright point-like or high-surface brightness 
extended sources was dictated by the nature of our detection algorithm, 
which leads to an excessive number of spurious detections at the 
periphery of the bright source, which is a common problem to most detection algorithms.
Therefore, each field was visually inspected to check for such effects
(see Figure~\ref{data});
when found, a conservatively large portion of the field was flagged. 
Of the 147 fields analysed, 11 ($\sim 7$\%) were discarded because of problems
at various stages of the pipeline execution. 
As a result of our selection, we retained 136 fields. 
Figure~\ref{aitoff_fluxes} (left) shows the Aitoff projection in 
Galactic coordinates of their positions. 
We note that several fields were observed more than once.
These fields were considered as different pointings, so that the number of 
distinct fields is 94.

The data input in our pipeline are  Level 2 (L2) data generated by the {\it Chandra} X-ray Center 
(CXC)  standard data processing in a uniform fashion,
which were filtered to only include the standard event grades, and
then corrected for aspect offsets. 
We applied energy filters to these event lists, 
and created soft (SB, 0.5--2.0\,keV), hard (HB, 2.0--7.0\,keV) and total 
(FB, 0.5--7.0\,keV) band event files. 
The upper limit on our hard and total energy bands was chosen at 
7\,keV because at higher energy the background increases and the effective 
area decreases, producing lower  signal-to-noise (S/N) data.
Our results in the 0.5--10\,keV band are then extrapolations 
from our findings in the 0.5--7\,keV range.

\subsection{The algorithm}

The main steps of the BMW algorithm can be summarised as follows 
(full details in \cite{Lazzatiea98}; \cite{Lazzatiea99};  \cite{Campanaea99}).
The first step is the creation of the WT of the input image; the BMW WT 
is based on the discrete multi-resolution theory and on the ``\`a{}
trous'' algorithm, 
which differs from continuous--WT-based algorithms which can 
sample more scales at the cost of a longer computing time. 
We used a Mexican hat mother-wavelet, which can be analytically approximated 
by the difference of two Gaussian functions. 
The WT decomposes each image into a set of sub-images, each of them carrying
the information of the original image at a given scale. This property  
makes the WT well suited for the analysis of X-ray images,
where the scale of sources is not constant over the field of view,
because of the dependence of the PSF on the off-axis angle.
We used 7 WT scales $a=[1,2,4,8,16,32,64]$ pixels, to cover a wide
range of source sizes, where $a$ is the scale of the WT
\cite{Lazzatiea99}.

Candidate sources are identified as local maxima above the significance threshold 
in the wavelet space at each scale, so that a list is obtained at each scale, 
and then a cross-match is performed among the 7 lists to merge them.
At the end of this step, we have a first estimate of source positions
(the pixel with the highest WT coefficient),  
source counts (the local maximum of the WT) 
and a guess of the source extension (the scale at which the WT is maximized). 
A critical parameter is the detection threshold which, in the context of WT algorithms,
is usually fixed arbitrarily by the user in terms of expected spurious detections 
per field \cite{Lazzatiea98}. 
The number of expected spurious detections as a function of the threshold 
value and for each scale was calculated 
by means of Monte Carlo simulations \cite{Morettiea02}.

The final step is the characterisation of the sources by means of a multi-source 
$\chi ^2$ minimization with respect to a Gaussian model source in the WT space. 
In order to fit the model on a set of independent data, the WT coefficients are 
decimated according to a scheme described in full in \cite{Lazzatiea99}.

We ran the detection algorithm on the source images 
rebinned by a factor of 2 (1 pixel $\sim 0.98$''), 
and then in their inner $512\times512$ part at the full resolution,
using 7 scales. 
We applied corrections to the source counts for vignetting and 
PSF modelling (i.e.\ for using a Gaussian to approximate the PSF function 
to fit the sources in wavelet space). 
We excluded the $480\times480$ pixel central part in the analysis at rebin 2,
then cross-correlated the positions of the sources found at rebin 1 and 2 to 
exclude common double entries.
We repeated this procedure for each of the three energy bands, and  
cross-correlated the resulting source coordinates to form the definitive list
(for coincident sources, the coordinates of the highest S/N one were
kept). 
We ran the detection algorithm with a single significance threshold 
that corresponds to $\sim 0.1$ spurious detections per scale, hence (with 
7 scales) $\sim 0.7$ spurious detections per field for each band in which we performed the detection. 
Given our sample of 136 fields, we expect a total of $\sim 95$ spurious 
sources in the catalogue, or a percentage of 2.7\% (considering the
three energy bands, the two images over which the detection was run,
and the total number of detected sources, see Sect.~\ref{catalogue}). 
An example of the results of the detection is shown in 
Figure~\ref{data}.

\subsection{The catalogue\label{catalogue}}

\begin{figure}
\includegraphics[width=0.5\textwidth,angle=0]{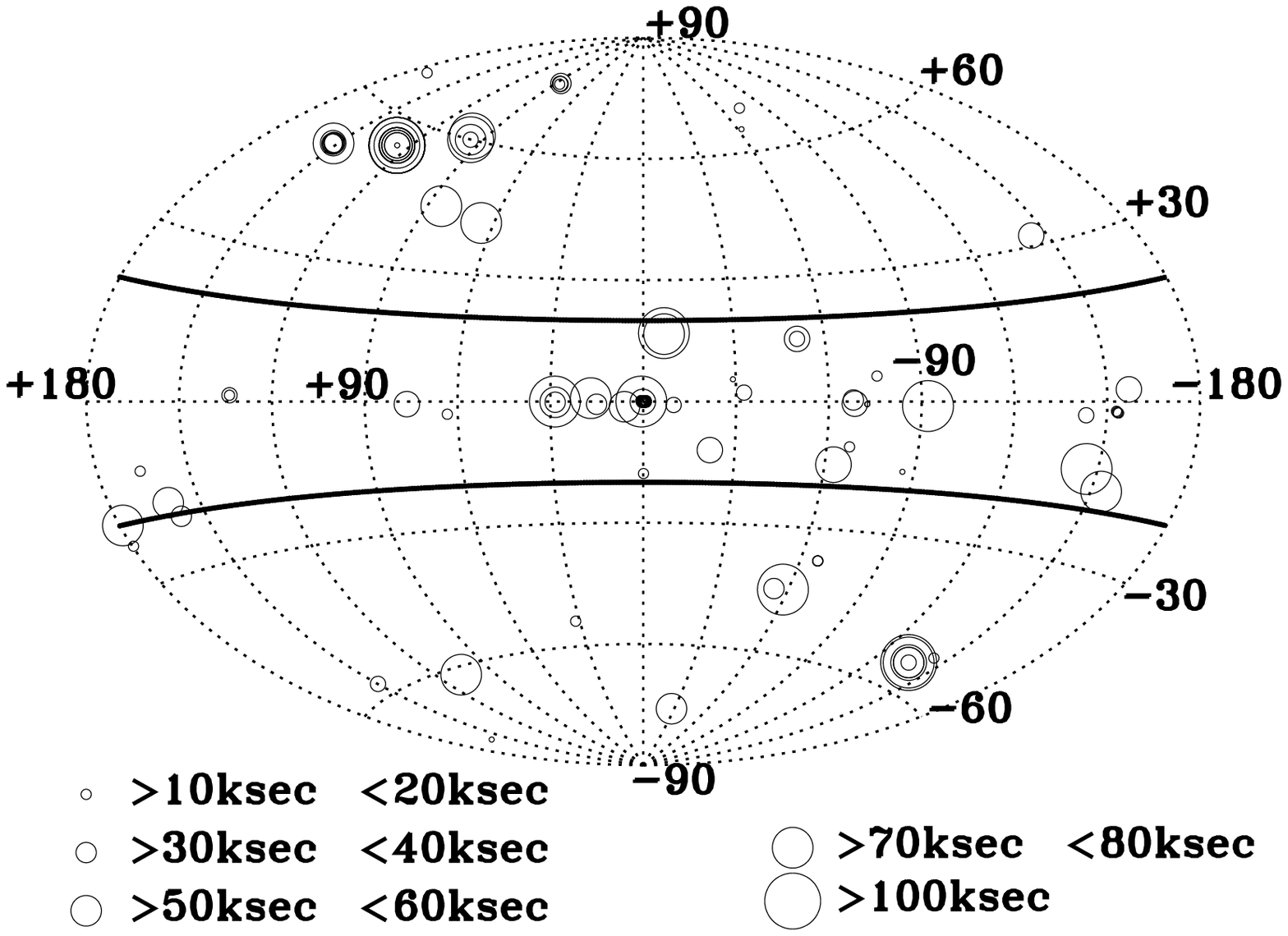}
\includegraphics[width=0.5\textwidth,angle=0]{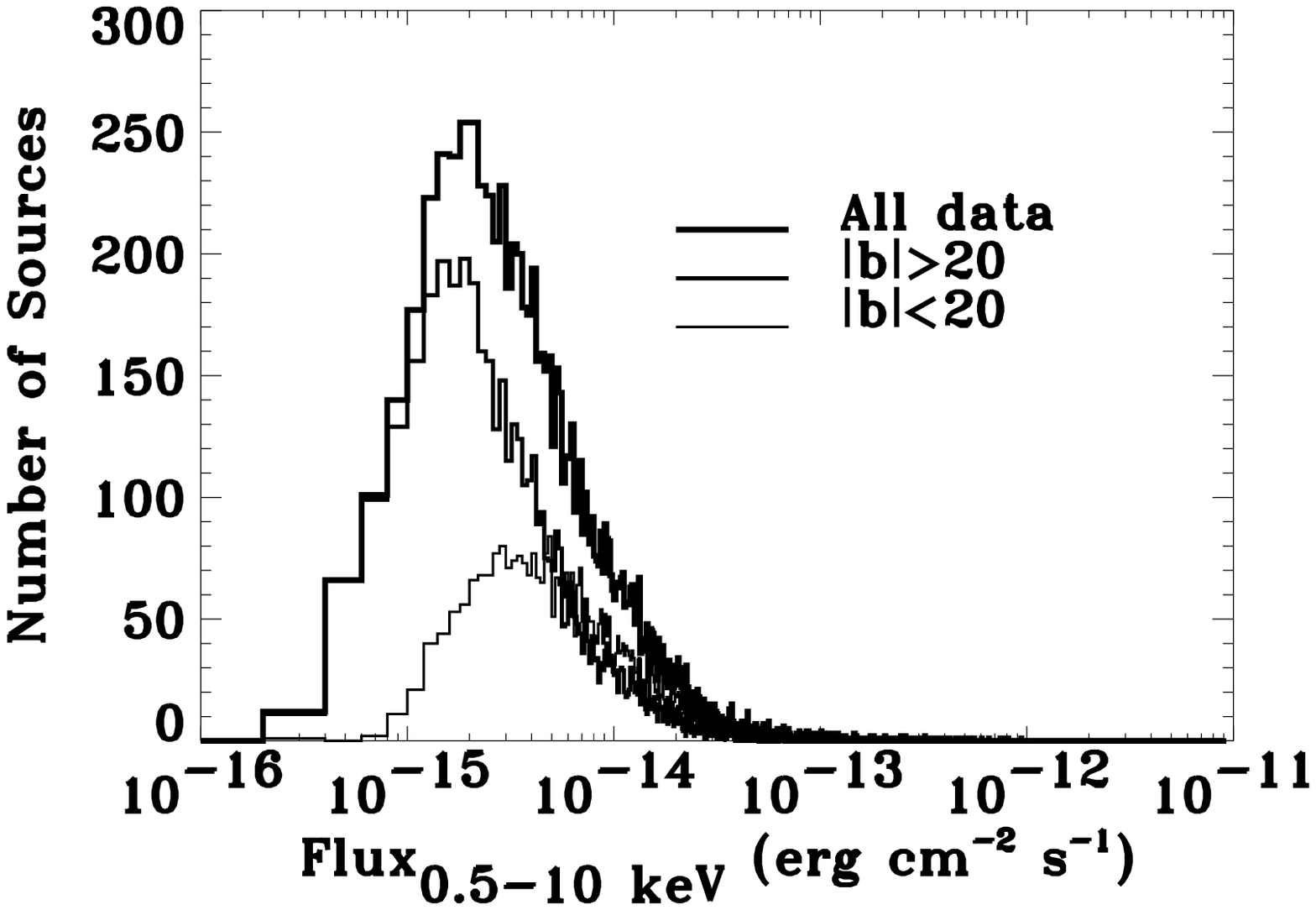}
\caption{{\bf Left}: Aitoff Projection in Galactic coordinates of the selected 136
  Chandra ACIS-I fields.  The thick lines are the limits for the high latitude sub-sample.
{\bf Right}: Distribution of the absorption
corrected 0.5--10\,keV flux in the full sample,
high latitude sample (7401), and low latitude
sample (9433).
}
\label{aitoff_fluxes}
\end{figure}

The wavelet detection produced a catalogue of source positions, count rates, 
counts, extensions, and relative errors in three bands, as well as the additional 
information drawn from the headers of the original files for a total of 21325 sources. 
We also extracted source counts within a box centered around the positions 
determined with the detection algorithm, with a side which is 
the 90\% encircled energy diameter at 1.50\,keV.
For the SB, HB, and FB bands the background counts were 
extracted from the same box from the background image.
We extracted source counts in the four additional bands: 
SB1 (0.5--1.0\,keV), 
SB2 (1.0--2.0\,keV), 
HB1 (2.0--4.0\,keV), and 
HB2 (4.0--7.0\,keV). 
We calculated the 0.5--10\,keV absorption corrected fluxes
by converting the count rates in fluxes assuming a Crab spectrum, 
i.e.\ a power law with photon index 2.0, modified with the absorption by 
Galactic $N_{\rm H}$ relative to each field; we also provide the 0.5--10\,keV observed
fluxes (simple Crab spectrum and $N_{\rm H}=0$). 
The catalogue lists the 0.5--10\,keV observed flux, the absorption corrected one and the 
corresponding conversion factors [see Figure~\ref{aitoff_fluxes} (right)]. 
Problematic portions (such as extended pointed objects) and pointed objects 
(within a radius of 30 arcsec from the target position) were flagged.

\begin{figure}
\includegraphics[width=0.5\textwidth,angle=0]{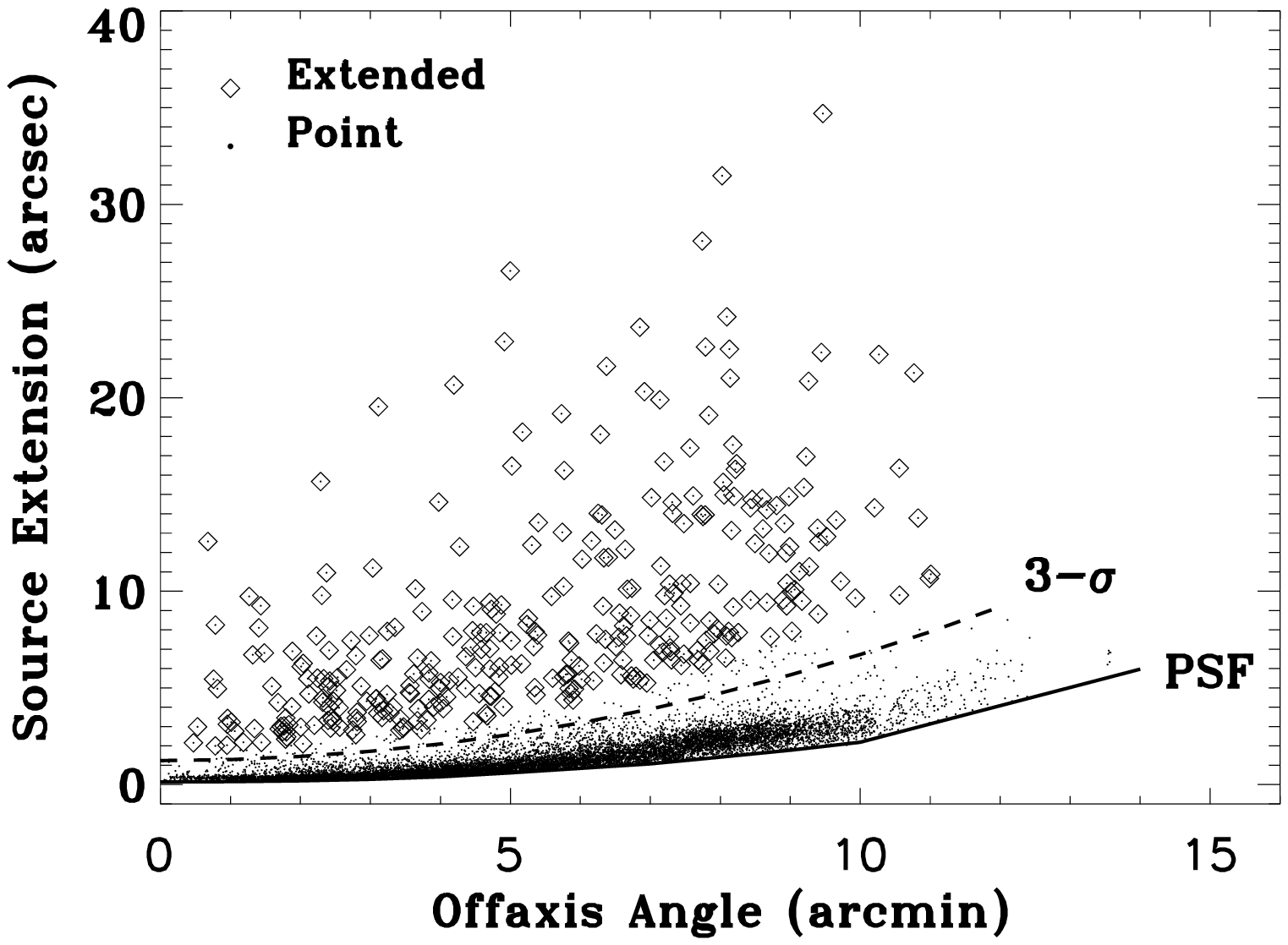}
\includegraphics[width=0.5\textwidth,angle=0]{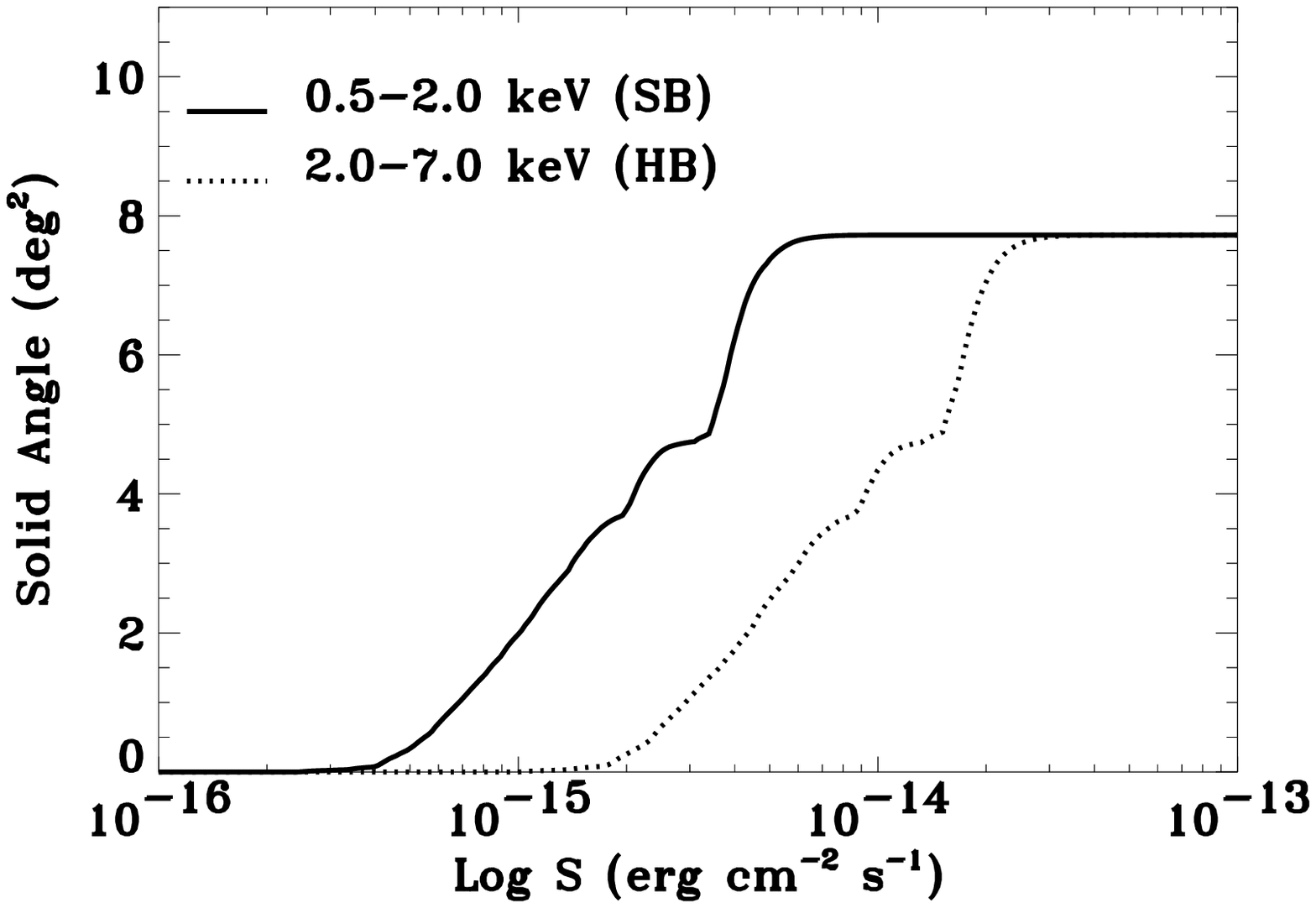}
\caption{{\bf Left}: Extension of the BMW {\it Chandra} sources as a function of off-axis angle. The solid line is the
PSF function, the dashed line is the 3$\sigma$ limit for point sources.
{\bf Right}: Solid angle versus flux limit for S/N $=3$ for the 
	soft (solid line) and hard (dotted line) bands. This sky coverage 
	was constructed using 94 independent fields (no fields covered the same sky area).}
\label{extension_skycoverage}
\end{figure}

Figure~\ref{extension_skycoverage} (left) shows the distribution of the source off-axis angle, 
which presents a steep increase with collecting area, and a gentler decrease with 
decreasing sensitivity with off-axis angles. Differently from what found with the 
BMW-HRI catalogue \cite{Panzeraea03}, our distribution does not present 
a peak at zero off-axis due to pointed sources. 
To characterise the source extension, 
which is one of the main features of the WT method, one cannot simply compare 
the WT width with the instrumental PSF at a given off-axis angle. 
Thus, we use a $\sigma$-clipping algorithm which divides the 
distribution of source extensions as a function of off-axis angle 
in bins of 1' width. 
The mean and standard deviation are calculated within each bin and all sources 
which width exceeds 3$\sigma$ the mean value are discarded. 
The procedure is repeated until convergence is 
reached. The advantage of this method is that it effectively eliminates truly 
extended sources, while providing a value for the mean and standard deviation 
in each bin \cite{Lazzatiea99}. The mean value plus the 3$\sigma$
dispersion 
provides the line discriminating the source extension, but we conservatively classify as 
extended only the sources that lie 2$\sigma$ above this limit. 
Combining this threshold with the 3$\sigma$ on the intrinsic dispersion, 
we obtain a $\sim 4.5$$\sigma$ confidence level for the extension classification.

The full catalogue contains 21325 sources, 16834 of which are 
not associated with bright and/or extended sources, including the
pointed ones. Of these, 11124 are detections in the total band,
12631 in the soft, 9775 in the hard band; 4203 sources were only detected in the hard
band (see Table~\ref{bmwc:numeri}).

It is particularly important for cosmological studies to have a sample which is 
not biased toward bright objects. To this end, we constructed the 
{\bf BMW-{\it Chandra} Serendipitous Source Catalogue that contains 16758 
sources not associated with pointed objects}, by 
excluding sources within a radius of 30 arcsec from the target position. 
Their sky coverage is shown in Figure~\ref{extension_skycoverage} (right).

\section{Catalogue Exploitation}

Among the avenues of scientific exploitation are: \\
1) Characterization of the sources based on X-ray colours alone; \\
2) Cross-correlation with other catalogues (FIRST, IRAS, 2MASS, GSC2)
allowed the identification of radio-to-optical counterparts; 
sub-samples of promising sources for optical follow-up include:

{\it i)} blank fields (sources without counterparts at other
wavelengths; Mignani et al.\ in prep.);

{\it ii)} heavily absorbed sources (the 4203 only detected in the hard X-ray
band); \\
3) Analysis of a sample of ~300 extended sources (Fig.~\ref{extension_skycoverage}), which constitutes
a list of X-ray selected galaxy cluster candidates, to confirm
optically (Romano et al. in prep.); \\
4) Temporal and spectral variability:
 
{\it i)} autocorrelation of the catalogue allows study of long-term variability
of sources observed more than once (Israel et al.\ in prep.); 

{\it ii)} intra-observation variability: search for periodicities in the light curves.

\section{BMW-{\it Chandra} online}

The current version of the BMW-{\it Chandra} source catalogue,
(as well as additional information and data) is available at the
Brera Observatory and at the  INAF-IASF
Palermo mirror sites, 
\begin{verbatim}
http://www.brera.inaf.it/BMC/bmc_home.html
http://www.ifc.inaf.it/~romano/BMC/bmc_home.html 
\end{verbatim}
The distributed version can also be found at the Centre de Donn\'ees
astronomiques de Strasbourg (Vizier) and at the HEASARC sites. 

\begin{table}
\begin{center}
\caption{BMW-C in short.}
\label{bmwc:numeri}
\begin{tabular}{llr}
\hline
\hline
\noalign{\smallskip}
Source Sample & & 				Number \\
\noalign{\smallskip}
\hline
\noalign{\smallskip}
  detected    			&    				&    	21325  \\
  good$^{\mathrm{a}}$ 		&    				&    	16834  \\
  serendipitous 	 	&				& 	16758  \\
  independent 			& (within 3'')	                &  	12135 \\
				& (within 4''.5)                & 	11954 \\
  detected in total band	& 				& 	11124 \\
  detected in soft band		&				& 	12631 \\
  detected in hard band		&				& 	9775 \\
  only detected in hard band	&				& 	4203 \\
  serendipitous extended 	&    				&    	316  \\ 
\noalign{\smallskip}
\hline
\end{tabular}
\end{center}
\begin{list}{}{}
\item[$^{\mathrm{a}}$] Sources which do not require a more in-depth, non-automated analysis 
(i.e.\ not associated with bright and/or extended sources at the centre of the field), 
including the target ones.
\end{list}
\end{table}

\acknowledgments

This work was supported through Consorzio Nazionale per l'Astronomia
e l'Astrofisica (CNAA) and Ministero dell'Istruzione, 
dell'Universit\`a{} e della Ricerca (MIUR) grants. 
We thank A.\ Mist\`o{} for his help with the database software. 
RPM acknowledges STFC for support through its Rolling Grant
programme. 
This publication makes use of data products from the {\it Chandra} Data Archive, 
the FIRST, IRAS, 2MASS, GSC2 surveys.

\end{document}